\documentclass[a4paper,twocolumn]{article}
\usepackage{nolta2012}
\usepackage{txfonts}
\usepackage{epsfig}
\usepackage{subfigure}
\usepackage{color}
\usepackage{setspace}
\usepackage{graphicx}
\usepackage{dcolumn}
\usepackage{bm}
\input epsf

\graphicspath{{figures/}}

\begin{document}

\title{Synchronization of Kuramoto oscillators in networks of networks}

\author{Per Sebastian Skardal and Juan G. Restrepo}
\address{Department of Applied Mathematics, University of Colorado at Boulder, UCB 526, \\
Boulder, Colorado 80309-0526, USA\\
Email: skardal@colorado.edu, juanga@colorado.edu
}

\maketitle

\abstract
We study synchronization of Kuramoto oscillators in strongly modular networks in which the structure of the network inside each community is averaged. We find that the dynamics of the interacting communities can be described as an ensemble of coupled planar oscillators. In the limit of a large number of communities, we find a low dimensional description of the level of synchronization between the communities. In this limit, we describe bifurcations between incoherence, local synchrony, and global synchrony. We compare the predictions of this simplified model with simulations of heterogeneous networks in which the internal structure of each community is preserved and find excellent agreement. Finally, we investigate synchronization in networks where several layers of communities within communities may be present.
\endabstract

\section{Introduction}

Large networks of coupled oscillators are found in many applications of science and engineering, including synchronized flashing of fireflies~\cite{FireFly}, cardiac pacemaker cells~\cite{Pacemaker}, oscillations of pedestrian bridges~\cite{Millenium}, and circadian rhythms in mammals~\cite{Circadian}. The Kuramoto model~\cite{Kuramoto} has become a paradigm for modeling and studying emergence of collective behavior in the form of synchronization. In the Kuramoto model each oscillator is described by a phase angle $\theta_n$ that evolves as
$
\dot \theta_n = \omega_n + K\sum_{m=1}^N A_{nm} \sin(\theta_m - \theta_n),
$
where $\omega_n$ is the intrinsic frequency of oscillator $n$, $K$ is the global coupling strength, $A_{nm}$ encodes the network topology, and $n,m=1,\dots,N$.

The path to synchrony in typical non-modular networks is characterized by incoherence for small $K$, followed by the emergence of a single synchronized cluster when $K$ surpasses a critical value $K_c$~\cite{Restrepo2005PRE}. However, when the network structure is modular, synchrony can occur hierarchically: first locally within each community, and then globally as communities synchronize with one another. Several studies on synchronization in modular networks exist~\cite{Pikovsky2008PRL,Barreto2008PRE,Arenas2006PRL,Skardal2012PRE}, but few analytic results for large networks of heterogeneous oscillators with many communities exist. 

In this paper we study a system of equations previously explored in Refs.~\cite{Barreto2008PRE,Skardal2012PRE,Ott2008Chaos} using the dimensionality reduction techniques of Ott and Antonsen~\cite{Ott2008Chaos}. We find analytical expressions for local and global order parameters describing synchronization within communities and on the whole network, respectively, and completely characterize the phase space of the system.

This paper is organized as follows. In Sec.~2 we summarize the results from Ref.~\cite{Skardal2012PRE} where network structure within communities is averaged. In Sec.~3 we compare the results of the averaged system to numerical simulations where the structure of each community is preserved. In Sec.~4 we show that results from Sec.~2 generalize to systems with multiple levels of community structure by studying the case of three levels. In Sec.~5 we conclude with a brief discussion.

\section{Hierarchical Synchrony in Two-level Hierarchical Networks}

We consider a system of $C$ communities, labeled $\sigma = 1,\dots,C$, each containing $N_\sigma$ oscillators. Oscillator $n$ in community $\sigma$ has phase $\theta_n^\sigma$ and evolves according to~\cite{Barreto2008PRE,Ott2008Chaos}
\begin{equation}\label{eqModel}
\dot{\theta}_n^\sigma=\omega_n^\sigma + \sum_{\sigma'=1}^C\eta_{\sigma'}\frac{K^{\sigma\sigma'}}{N_{\sigma'}}\sum_{m=1}^{N_{\sigma'}}\sin(\theta_m^{\sigma'}-\theta_n^\sigma),
\end{equation}
where $\omega_n^\sigma$ is its intrinsic frequency, $K^{\sigma\sigma'}$ is the coupling between oscillators in communities $\sigma$ and $\sigma'$, and $\eta_\sigma$ is the fraction of oscillators in community $\sigma$. 

For analytic tractability, we make the following simplifications. First, we assume all communities are of the same size, i.e. $N_\sigma=N$ and $\eta_\sigma=C^{-1}$. In addition, we assume that (i) the coupling strength between oscillators within the same community is much larger than the coupling strength between oscillators in different communities and (ii) the intrinsic frequency for an oscillator is drawn from a distribution specific to its own community. To ensure condition (i) for a large number of communities, we let $K^{\sigma\sigma'}=Ck$ if $\sigma=\sigma'$, and $K$ otherwise, where $k$ and $K$ are of the same order. Finally, to ensure condition (ii) we assume that the frequency $\omega_n^\sigma$ is drawn from a distribution $g^\sigma(\omega)$. We will assume that this distribution is Lorentzian with uniform spread $\delta$ and community-specific mean $\Omega_\sigma$: $g_\sigma(\omega)=\pi^{-1}\delta/[\delta^2+(\omega-\Omega_\sigma)^2]$. Furthermore, the means $\Omega_\sigma$ are drawn from their own distribution $G(\Omega)$ which is Lorentzian with spread $\Delta$ and mean zero: $G(\Omega)=\pi^{-1}\Delta/(\Delta^2+\Omega^2)$. Parameters $\delta,\Delta=1$ are used for all figures presented. A discussion of how some of these assumptions can be relaxed can be found in Ref.~\cite{Skardal2012PRE}.

Finally, to characterize the degree of local and global synchrony, we define the complex order parameters
\begin{eqnarray}
z_{\sigma} = r_\sigma e^{i\psi_\sigma}=\frac{1}{N}\sum_{m=1}^{N}e^{i\theta_m^\sigma}, \hskip2ex
Z = Re^{i\Psi}= \frac{1}{C}\sum_{\sigma=1}^C z_\sigma, \label{eqOrdR}
\end{eqnarray}
such that $r_\sigma$ measures the degree of local synchrony in community $\sigma$ and $R$ measures the degree of global synchrony over the entire network. To measure the average degree of local synchrony we introduce $\overline{r} = \frac{1}{C}\sum_{\sigma}r_\sigma$.

\begin{figure}[t]
\addtolength{\belowcaptionskip}{-4mm}
\centering
\epsfig{file =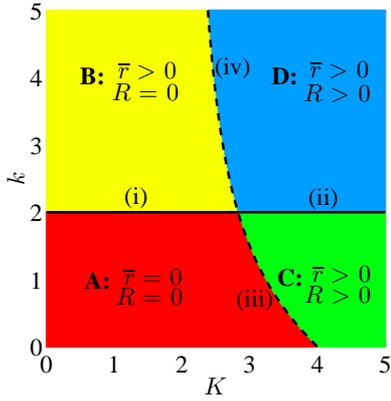, clip =,width=0.70\linewidth }
\caption{Phase space for Eq.~(\ref{eqModel}) with $\delta,\Delta=1$. Regions A, B, C, and D (described in the text) are denoted in red, yellow, green, and blue, respectively, with bifurcations (i)-(iv) indicated by solid and dashed curves.} \label{BifD}
\end{figure}

The phase space of this system as a function of $K,k$ was found in~\cite{Skardal2012PRE} in the limits $N,C\to\infty$ and is shown in Figure~\ref{BifD}. For small $K,k$ (A: red) both the average local degree of synchrony $\overline{r}$ and global degree of synchrony $R$ are zero. If $k$ is increased and $K$ is kept sufficiently small (B: yellow) local synchrony is non-zero but no global synchrony exists. Finally, for sufficiently large $K$ (C: green and D: blue) both local and global synchrony are non-zero. As we will discuss, these regions differ in how communities synchronize with one another. Bifurcation curves between these different states are labelled (i)--(iv) and will be discussed below. We now summarize some results from Ref.~\cite{Skardal2012PRE} on local and global synchrony.

\subsection{Local synchrony}

To classify local synchrony we consider the continuum limit $N\to\infty$ and introduce the disribution $f_\sigma(\theta,\omega,t)$ that describes the density of oscillators with phase and frequency $\theta$ and $\omega$ at time $t$. After a dimensionality reduction (described in detail in Refs.~\cite{Skardal2012PRE,Ott2008Chaos}), we find that $r_\sigma$ and $\psi_\sigma$ evolve according to the $2C$-dimensional planar oscillator system
\begin{eqnarray}
\dot{r}_\sigma = -r_\sigma\delta + \left(k-\frac{K}{C}\right)r_\sigma\frac{1-r_\sigma^2}{2}+K\frac{1-r_\sigma^2}{2}R\cos(\Psi-\psi_\sigma)\\
\dot{\psi}_\sigma = \Omega_\sigma+K\frac{1+r_\sigma^2}{2r_\sigma}R\sin(\Psi-\psi_\sigma). \hskip8ex
\end{eqnarray}

In regions A and B, where $R=0$, the communities decouple and the steady-state degree of local synchrony for each community is given by
\begin{equation}\label{eqrsigma}
r_\sigma=\left\{\begin{array}{ll} 0 &\mbox{ if }k-K/C \le 2\delta, \\ \sqrt{1-\frac{2\delta}{k-K/C}} & \mbox{ otherwise,}\end{array}\right.
\end{equation}
and the onset of local synchrony is given by bifurcation (i) $k-K/C=2\delta$.

Now we analyze regions C and D. Given a steady-state $R$ value (to be discussed later) it can be shown~\cite{Skardal2012PRE} that in region C all communities lock and their degree of local synchrony $r_\sigma$ is given implicitly by 
 \begin{eqnarray}\label{eqrhoimp}
r_\sigma\delta&=&\left(k-\frac{K}{C}\right)r_\sigma\frac{1-r_\sigma^2}{2} \nonumber \\ &\hskip2ex& +KR\frac{1-r_\sigma^2}{2}\sqrt{1-\frac{4\Omega_{\sigma}^2r_\sigma^2}{K^2R^2(r_\sigma^2+1)^2}}.
\end{eqnarray} 
In region D only a fraction of the the communities lock, which are precisely those communities with sufficiently small mean frequency (in magnitude) given by
\begin{equation}\label{Y} 
|\Omega_{\sigma}| \le \widetilde{\Omega} \equiv KR\left(1-\frac{\delta^2}{(k-\frac{K}{C}-\delta)^2}\right)^{-1/2},
\end{equation}
and have a degree of local synchrony given by Eq.~(\ref{eqrhoimp}). The drifting communities turn out to have solutions where $r_\sigma$ oscillates with approximate mean $\sqrt{1-2\delta/(k-K/C)}$. Bifurcation (ii) is given by $k-K/C=2\delta$, which can be found by sending $\widetilde{\Omega}\to\infty$ in Eq.~(\ref{Y}).

\subsection{Global synchrony}

To classify global synchrony we consider the continuum limit $C\to\infty$ and introduce the distribution $F(\psi,\Omega,r,t)$ that describes the density of communities with phase, mean frequency, and degree of local synchrony $\psi$, $\Omega$, and $r$ at time $t$. After a dimensionality reduction (described in detail in Ref.~\cite{Skardal2012PRE}), we find that $R$ evolves according to
\begin{equation}
\dot{R} = -\Delta+\frac{K}{4}R(1+\hat{r}^2)\left(1-\frac{R^2}{\hat{r}^2}\right),
\end{equation}
where $\hat{r}$ solves Eq.~(\ref{eqrhoimp}) for $\Omega=-i\Delta$, while $\dot{\Psi}=0$. The steady-state degree of global synchrony is given by
\begin{equation}\label{eqRss}
R=\left\{\begin{array}{ll}0 & \mbox{ if }K\le\frac{4\Delta}{\hat{r}^2+1}, \\ \hat{r}\sqrt{1-\frac{4\Delta}{K(\hat{r}^2+1)}} &\mbox{ otherwise,} \end{array}\right.
\end{equation}
which can be solved consistently with $\hat{r}$. Sending $\hat{r}\to\sqrt{4\Delta/K-1}^+$ yields an onset of global synchrony given by $k = \frac{\delta K}{K-2\Delta}-\frac{K}{2}$ corresponding to bifurcations (iii) and (iv).

\begin{figure}[t]
\addtolength{\belowcaptionskip}{-4mm}
\centering
\epsfig{file =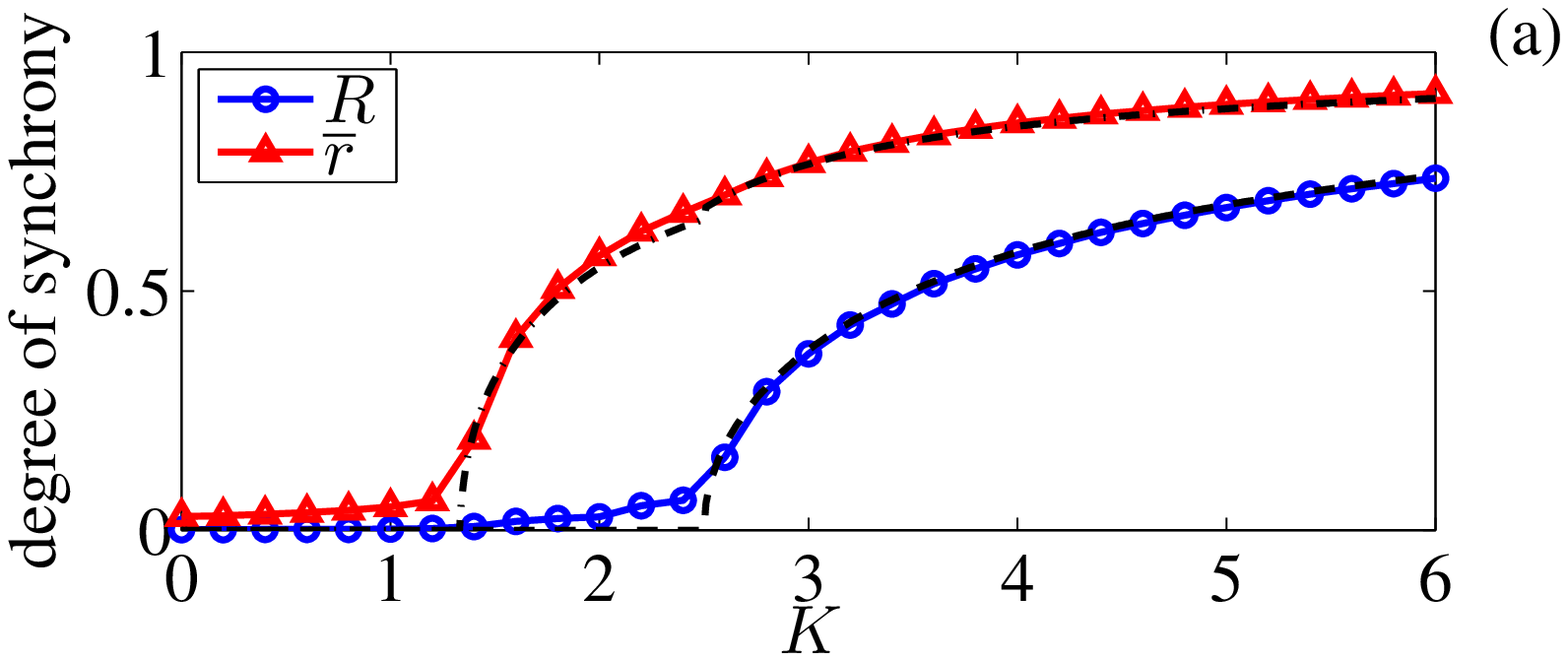, clip =,width=0.75\linewidth } \\
\epsfig{file =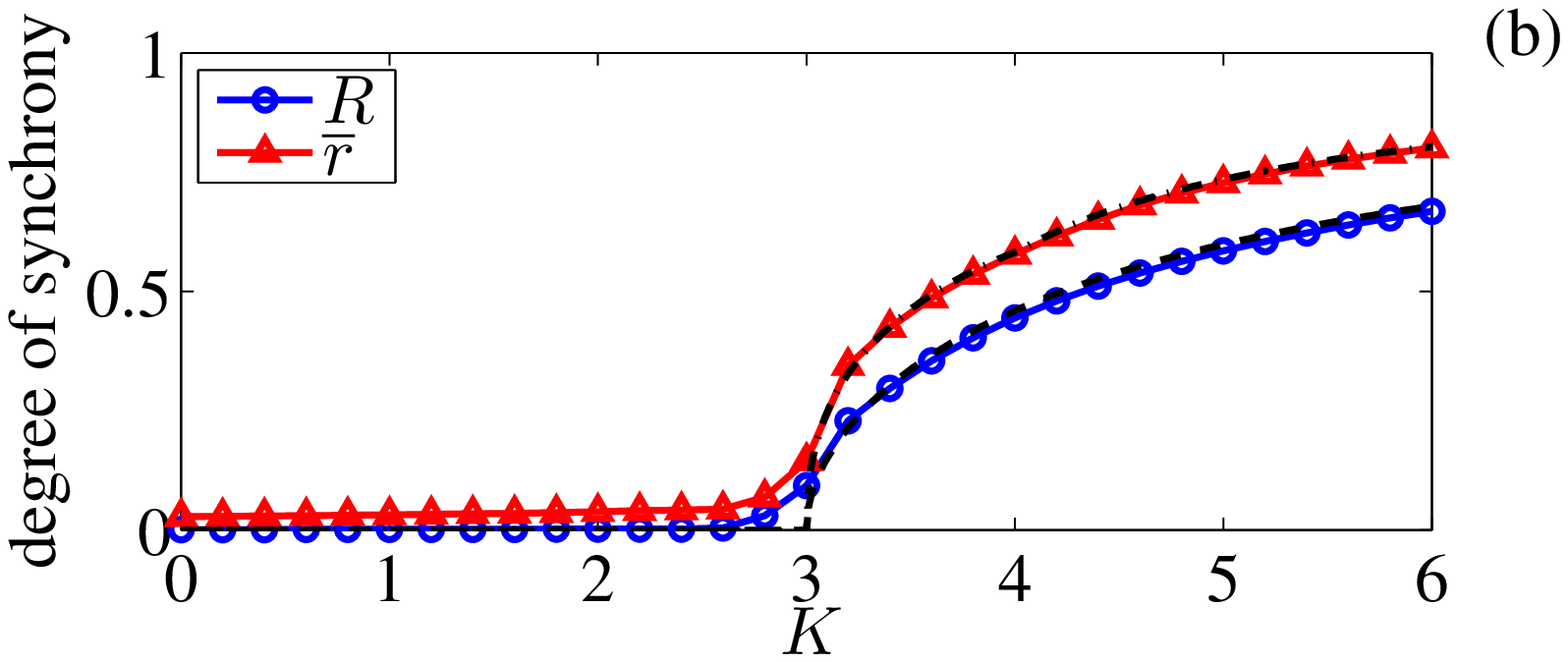, clip =,width=0.75\linewidth }
\caption{Degree of global and local synchrony (blue circles and red triangle) along paths (a) $k=3K/2$ and (b) $k=K/2$ from simulation with $N,C=1000$. Theoretical predictions are plotted in dashed black.} \label{hier}
\end{figure}

We now consider the implications of our analysis on the hierarchy of local and global synchrony. Starting in region A, the onset of local and global synchrony will occur separately if bifurcation (i) is reached first, otherwise bifurcation (iii) is reached first and they occur simultaneously. Given a path $k=mK$ where $K$ is increased from zero, if $m>m_c=\frac{2\delta}{\Delta-\delta+\sqrt{\Delta^2+\delta^2+6\delta\Delta}}$ then the relative ratio of local to global coupling is strong enough to produce hierarchical synchrony, where $\overline{r}>0$ before $R>0$. On the other hand, if $m<m_c$ global effects dominate at onset and $\overline{r}>0,R>0$ occur simultaneously. We plot the steady-state values of $R$ and $\overline{r}$ from simulations (blue circles and red triangles) and theory (dashed black curves) resulting from increasing $K$ while $k=mK$ for $N,C=1000$ in Fig.~\ref{hier} for $m=3/2$ and $1/2$ [subfigures (a) and (b), respectively]. We observe a hierarchical separation of the onset of local and global synchrony for $m=3/2$ but not for $m=1/2$.

\section{Numerical Experiments}
In this Section, we test the usefulness of the theoretical results in Section 2 in predicting behavior of non-trivial networks. We consider a system given by 
\begin{equation}\label{eqER}
\dot{\theta}_n^\sigma = \omega_n^\sigma + \frac{\widetilde{K}}{N}\sum_{\sigma'=1}^C\sum_{m=1}^NA_{nm}^{\sigma\sigma'}\sin(\theta_m-\theta_n),
\end{equation}
where $A_{nm}^{\sigma\sigma'}$ encodes the network structure between oscillators $n$ and $m$ in communities $\sigma$ and $\sigma'$, respectively, and $\widetilde{K}$ is the global coupling strength. We consider the case where the network in community $\sigma$ is an Erd\H{o}s-R\'{e}nyi random network. In order to attain modularity, we assume that when $\sigma=\sigma'$, $A_{nm}^{\sigma\sigma'}$ is 1 with probability $p_1\sim1$ and 0 otherwise, and that when $\sigma\ne\sigma'$, $A_{nm}^{\sigma\sigma'}$ is 1 with probability $p_2\ll1$ and 0 otherwise. Using the mean degrees $\langle d\rangle_{1,2}=Np_{1,2}$ we can estimate effective $k,K$ values as $k=\langle d\rangle_1\widetilde{K}/N=\widetilde{K}p_1$ and $K = C\langle d\rangle_2\widetilde{K}/N=C\widetilde{K}p_2$.

\begin{figure}[t]
\addtolength{\belowcaptionskip}{-4mm}
\centering
\epsfig{file =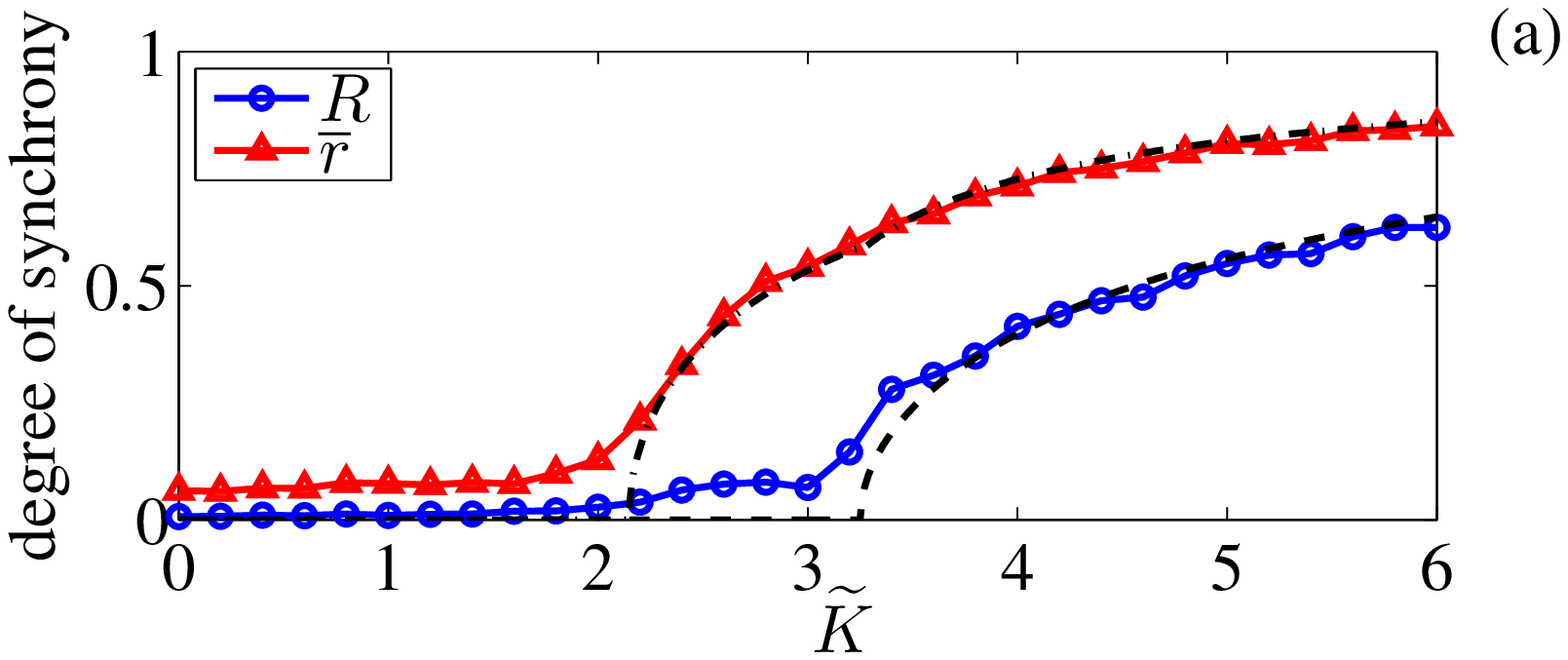, clip =,width=0.75\linewidth } \\
\epsfig{file =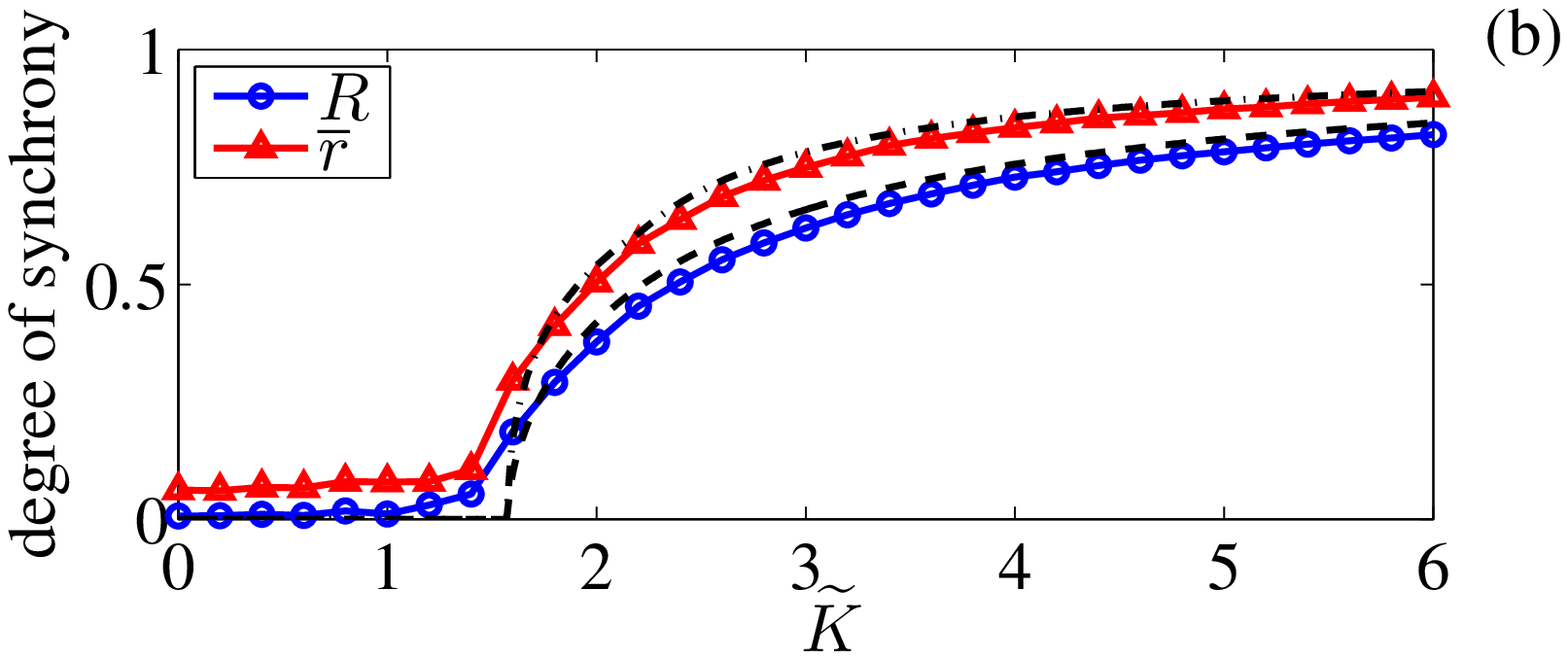, clip =,width=0.75\linewidth }
\caption{Degree of global and local synchrony (blue circles and red triangle) for Erd\H{o}s-R\'{e}nyi network simulations with $N=200$, $C=50$, $p_1=0.95$, and $p_2 = 0.012\overline{6}$ (a) and $0.038$ (b), respectively. Theoretical predictions from the averaged system are plotted in dashed black.} \label{ErdosRenyi}
\end{figure}

In Fig.~\ref{ErdosRenyi} we plot the degree of global and local synchrony from simulating Eq.~(\ref{eqER}) on Erd\H{o}s-R\'{e}nyi networks (blue circles and red triangles) compared to theoretical predictions from the community-averaged case (dashed black) using the corresponding values of $k$ and $K$ estimated in the previous paragraph. Parameters are $N = 200$, $C= 50$, $p_1=0.95$, and $p_2=0.012\overline{6}$ and $0.038$ [subplots (a) and (b), respectively]. The agreement between the theoretical prediction from the community-averaged system and actual simulation of the Erd\H{o}s-R\'{e}nyi networks is excellent.

\section{Hierarchical Synchrony in Multiple-Levels}

\begin{figure*}[t]
\addtolength{\belowcaptionskip}{-4mm}
\centering
\epsfig{file =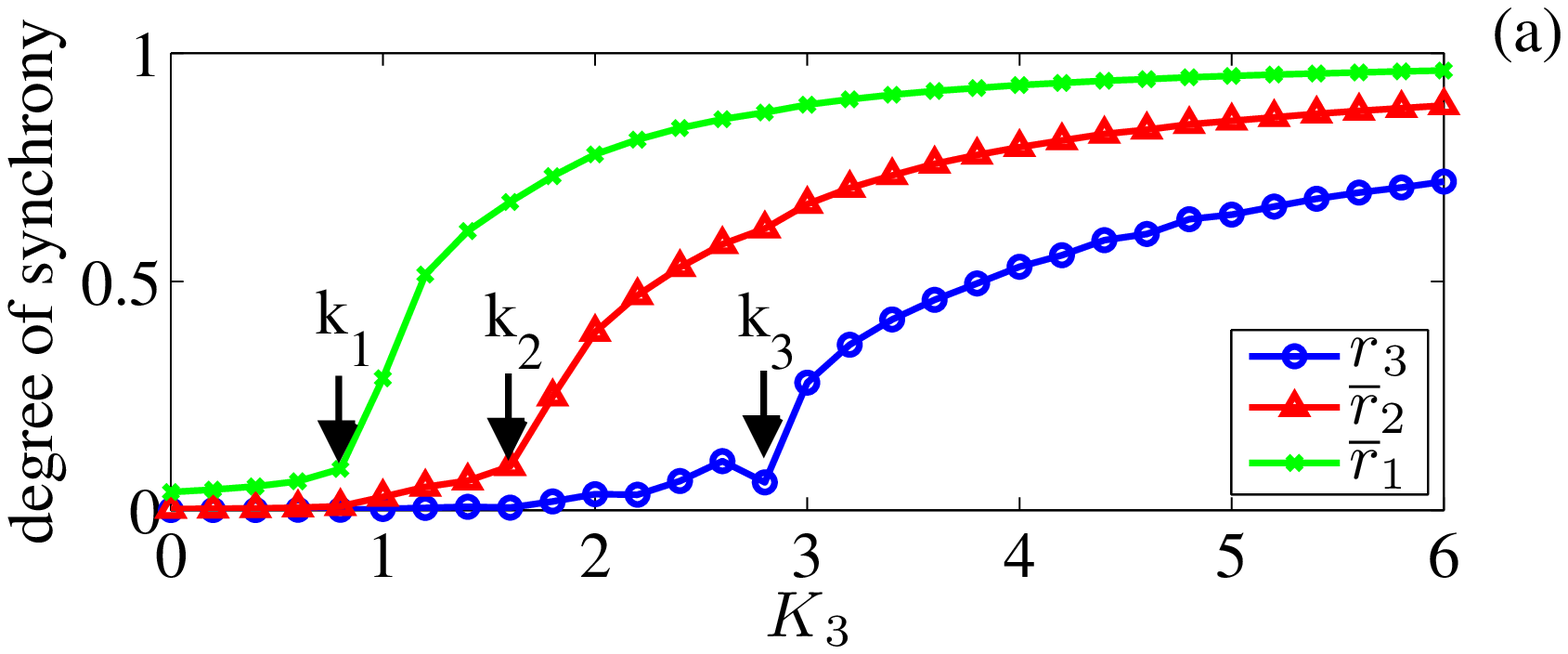, clip =,width=0.375\linewidth } \hskip5ex
\epsfig{file =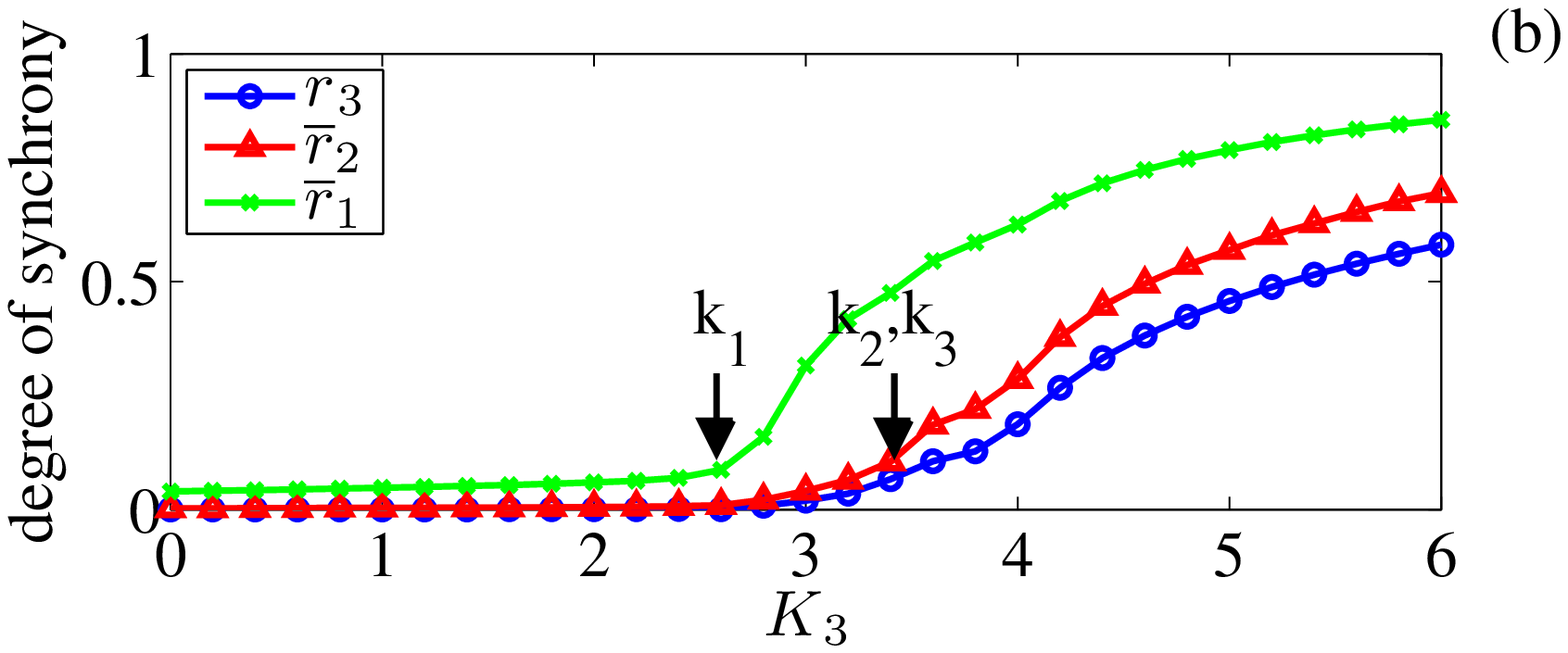, clip =,width=0.375\linewidth } \\
\epsfig{file =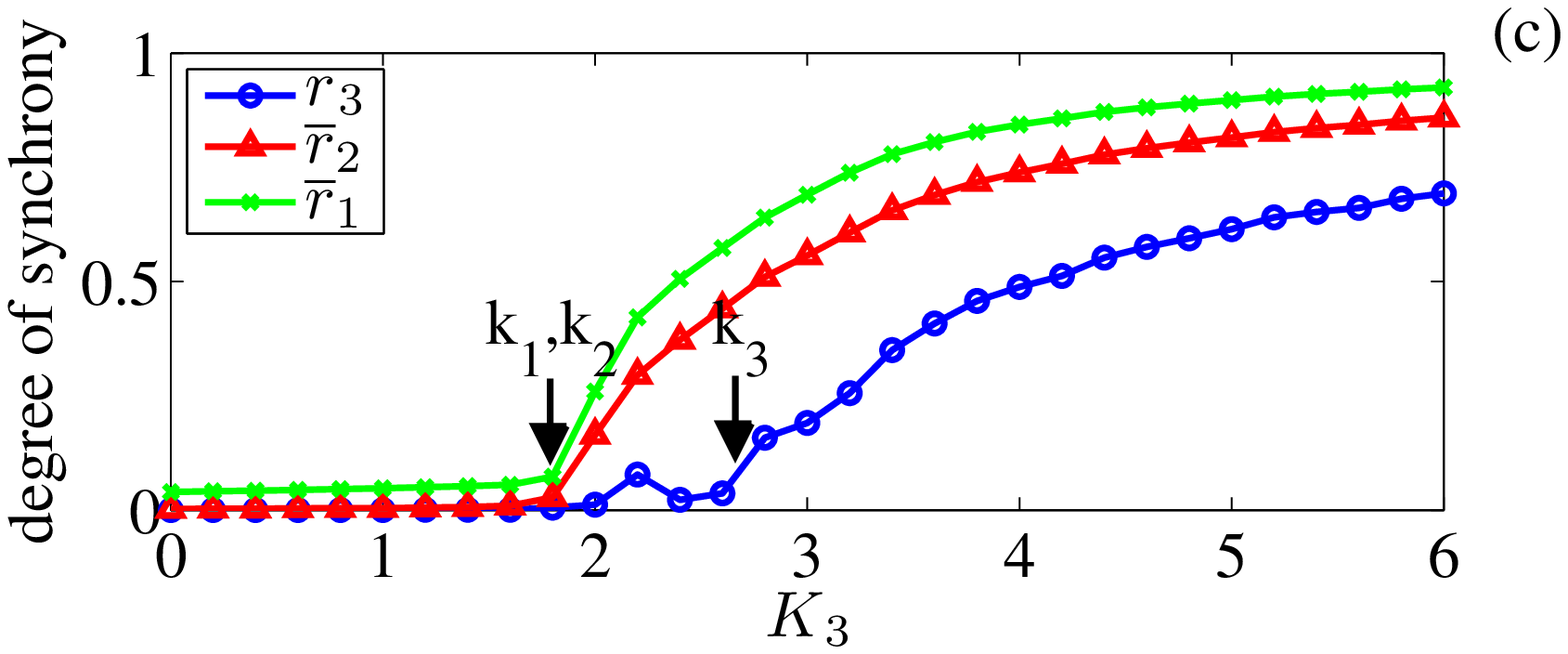, clip =,width=0.375\linewidth } \hskip5ex
\epsfig{file =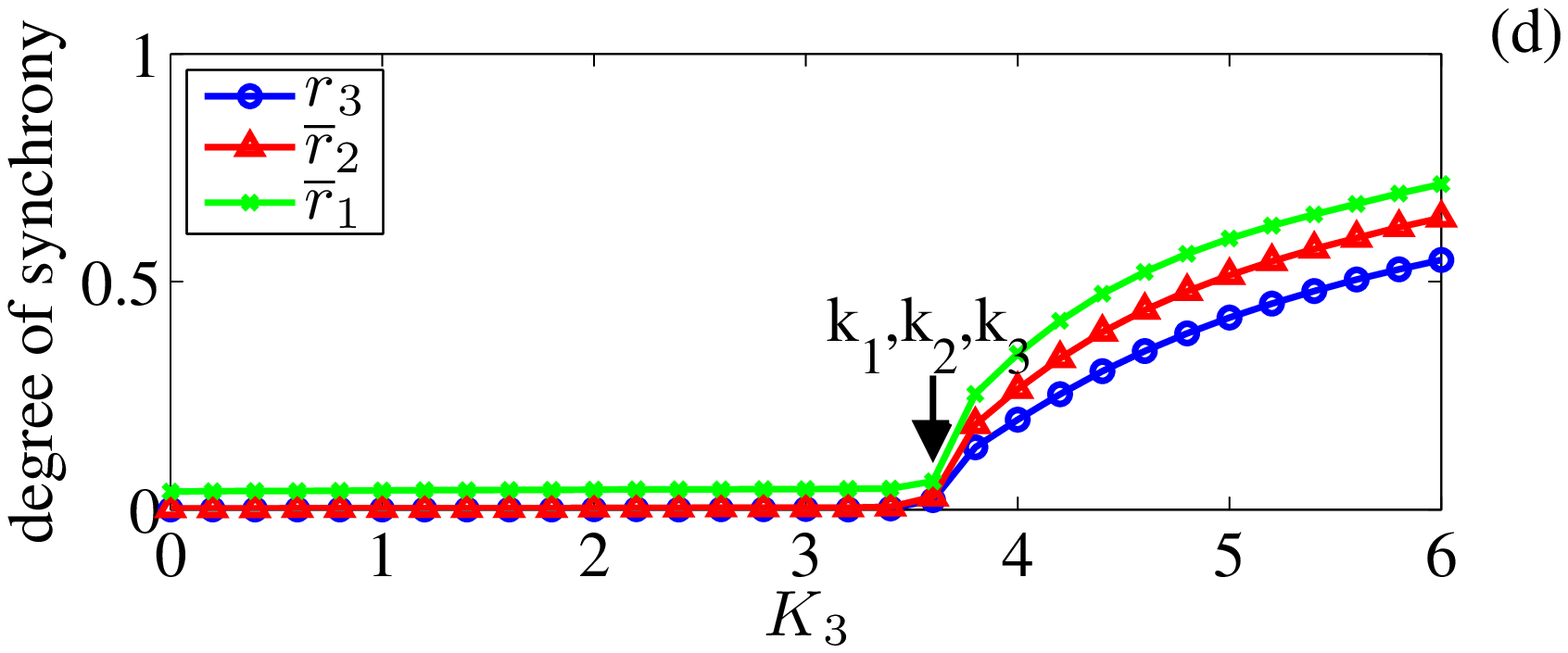, clip =,width=0.375\linewidth }
\caption{Degree of global, mid-level, and local synchrony (blue circles, red triangle, and green crosses) for three community levels with $N=500$, $S=200$, $C=120$. Coupling is described by paths $K_2=aK_3$, $K_1=bK_2$ for $a,b=\frac{3}{2},\frac{3}{2}$ (a), $\frac{3}{2},\frac{1}{2}$ (b), $\frac{1}{2},\frac{3}{2}$ (c), and $\frac{1}{2},\frac{1}{2}$ (d).} \label{three}
\end{figure*}

We now consider a system with three levels of community structure: a network of oscillators with $C$ communities where each community contains $S$ sub-communities, and each sub-community contains $N$ oscillators that evolve according to 
$$
\dot{\theta}_{n,\gamma,\sigma} = \omega_{n,\gamma,\sigma} + \frac{1}{C}\sum_{\sigma'=1}^C\frac{1}{S}\sum_{\gamma'=1}^S\frac{K_{\gamma\gamma'}^{\sigma\sigma'}}{N}\sum_{m=1}^N\sin(\theta_{m,\gamma',\sigma'}-\theta_{n,\gamma,\sigma}), \label{eqthree}
$$
where $\theta_{n,\gamma,\sigma}$ and $\omega_{n,\gamma,\sigma}$ denote the phase and frequency of oscillator $n$ in sub-community $\gamma$ of community $\sigma$ and $K_{\gamma\gamma'}^{\sigma\sigma'}$ denotes the coupling strength between oscillators in sub-community $\gamma$ of community $\sigma$ and sub-community $\gamma'$ in community $\sigma'$. To ensure multi-level community structure we assume that $K_{\gamma\gamma'}^{\sigma\sigma'}=SCK_1$ for $\gamma=\gamma',\sigma=\sigma'$, $CK_2$ for $\gamma\ne\gamma',\sigma=\sigma'$, and $K_3$ otherwise. In analogy with the two-level system, the frequencies $\omega_{n,\gamma,\sigma}$ are drawn from the Lorentzian $g_\gamma^\sigma(\omega)=\pi^{-1}\delta/[\delta^2+(\omega-\Omega_1^{\gamma\sigma})^2]$, $\Omega_1^{\gamma\sigma}$ is drawn from the Lorentzian $G_\sigma(\Omega)=\pi^{-1}\Delta/[\Delta^2+(\Omega-\Omega_2^\sigma)^2]$, and $\Omega_2^\sigma$ is drawn from the Lorentzian $\widetilde{G}(\Omega)=\pi^{-1}\widetilde{\Delta}/(\widetilde{\Delta}^2+\widetilde{\Omega}^2)$. To measure the degrees of local, mid-level, and global synchrony, we use the order parameters
$$
z_1^{\gamma\sigma}=\frac{1}{N}\sum_{n=1}^Ne^{i\theta_{n,\gamma,\sigma}}, \hskip2ex z_2^{\sigma} = \frac{1}{S}\sum_{\gamma=1}^Sz_1^{\gamma\sigma}, \hskip2ex z_3=\frac{1}{C}\sum_{\sigma=1}^Cz_2^\sigma,
$$
whose magnitudes $r_1^{\gamma\sigma}$, $r_2^\sigma$, and $r_3$ measure the degrees of synchrony and have phases $\psi_1^{\gamma\sigma}$, $\psi_2^\sigma$, and $\psi_3$. We measure the average local and mid-level degrees of synchrony by $\overline{r}_1=\frac{1}{SC} \sum_{\gamma,\sigma}r_1^{\gamma\sigma}$ and $\overline{r}_2=\frac{1}{C}\sum_{\sigma} r_2^\sigma$.

We investigate the hierarchy of local, mid-level, and global synchrony by moving through paths $K_2=aK_3$, $K_1=bK_2$ as we increase $K_3$ from zero. In Fig.~\ref{three} we plot the resulting $r_3$, $\overline{r}_2$, and $\overline{r}_1$ (blue circles, red triangles, and green crosses, respectively) for $a,b=\frac{3}{2},\frac{3}{2}$ (a), $\frac{3}{2},\frac{1}{2}$ (b), $\frac{1}{2},\frac{3}{2}$ (c), and $\frac{1}{2},\frac{1}{2}$ (d). Onsets of local, mid-level, and global synchrony, denoted $k_1$, $k_2$, and $k_3$, respectively, are indicated by arrows. For $a,b=\frac{3}{2},\frac{3}{2}$ the onsets of each level of synchrony are separated, while for $a,b=\frac{1}{2},\frac{1}{2}$ all occur simultaneously. For $a,b=\frac{3}{2},\frac{1}{2}$ mid-level and global synchrony occur simultaneously and are separated from local onset while for $a,b=\frac{1}{2},\frac{3}{2}$ local and mid-level synchrony occur simultaneously and are separated from global onset.

\section{Discussion}

We have presented analytical results describing local and global synchrony in modular networks where the network structure within each community is averaged. Furthermore, we have shown via numerical simulations that these analytical results predict very well the dynamics of non-trivial networks with community-wise Erd\H{o}s-R\'{e}nyi topologies. Importantly, analytical results indicate whether the path to synchrony occurs hierarchically or not. The effect of stronger heterogeneity on hierarchical synchrony in modular networks remains an open area of research. Finally, we have investigated hierarchical synchrony in networks with several layers of community structure and have found that depending on the relative ratios of the coupling strengths, synchrony of different layers can occur either hierarchically or simultaneously as in the two-layer case.

\section*{Acknowledgments}
The authors would like to thank the 2012 NOLTA organizing committee members for their invitation to participate in the conference. The work of P.S.S. and J.G.R. was supported by NSF Grant No. DMS-0908221.


\begin{thebibliography}{9}\setlength{\itemsep}{-2mm}
\bibitem{FireFly} J. Buck, ``Synchronous rhythmic flashing of fireflies,'' \textit{Q. Rev. Biol.} vol.63, pp.265--289, 1988.
\bibitem{Pacemaker} L. Glass and M. C. Mackey, \textit{From Clocks to Chaos: The Rhythms of Life} (Princeton University Press, Princeton, 1988).
\bibitem{Millenium} S. H. Strogatz et al., ``Crowd synchrony on the Millenium Bridge,'' \textit{Nature (London)} vol.438, pp.43--44, 2005.
\bibitem{Circadian} S. Yamaguchi et al., ``Synchronization of cellular clocks in the suprachiasmatic nucleus,'' \textit{Science} vol.302, pp.1408--1412, 2003.
\bibitem{Kuramoto} Y. Kuramoto, \textit{Chemical Oscillations, Waves, and Turbulence} (Springer, New York, 1984).
\bibitem{Restrepo2005PRE} J. G. Restrepo, E. Ott, and B. R. Hunt, ``Onset of synchronization in large networks of coupled oscillators,'' \textit{Phys. Rev. E} vol.71, 036151(1)--036151(12), 2005.
\bibitem{Pikovsky2008PRL} A. Pikovsky and M. Rosenblum, ``Partially integrable dynamics of hierarchical populations of coupled oscillators,'' \textit{Phys. Rev. Lett.} vol.101, pp.264103(1)--264103(4), 2008.
\bibitem{Barreto2008PRE} E. Barreto et al., ``Synchronization in networks of networks: The onset of coherent collective behavior in systems of interacting populations of heterogeneous oscillators,'' \textit{Phys. Rev. E}, vol.77, pp.036107(1)--036107(7), 2008.
\bibitem{Arenas2006PRL} A. Arenas, A. Diaz-Guilera, and C. J. Perez-Vicente, ``Synchronization reveals topological scales in complex networks,'' \textit{Phys. Rev. Lett.} vol.96, pp.114102(1)--114102(4), 2006.
\bibitem{Skardal2012PRE} P. S. Skardal and J. G. Restrepo, ``Hierarchical synchrony of phase oscillators in modular networks,'' \textit{Phys. Rev. E.}, vol.85, pp.016208(1)--016208(8), 2012.
\bibitem{Ott2008Chaos} E. Ott and T. M. Antonsen, ``Low dimensional behavior of large systems of globally coupled oscillators,'' \textit{Chaos} vol.18, pp.037113(1)--037113(6), 2008; ``Long time evolution of phase oscillator systems,'' \textit{Chaos} vol.19, pp.023117(1)--023117(5), 2009.
\end{thebibliography}
\end{document}